\documentclass{aastex}
\usepackage{emulateapj5,apjfonts,psfig}

\def\cm{{\rm\thinspace cm}}
\def\erg{{\rm\thinspace erg}}

\def\km{{\rm\thinspace km}}

\def\s{{\rm\thinspace s}}

\def\ergps{\hbox{$\erg\s^{-1}\,$}}
\def\kmps{\hbox{$\km\s^{-1}\,$}}

\def\psqcm{\hbox{$\cm^{-2}\,$}}

\def\etal{{et al.\ }}
\def\mcg6{MCG$-$6-30-15}
\def\chandra{{\it Chandra }}
\def\xmm{{\it XMM-Newton }}
\def\rxte{{\it RXTE }}
\def\asca{{\it ASCA }}

\def\xtegammamcg6{$\Gamma=1.9$}

\def\oviitau{$\tau_{\rm OVII}$ }

\def\oviiitau{$\tau_{\rm OVIII}$ }
\def\ovii{O~{\sc vii }}
\def\oviii{O~{\sc viii }}

\def\lapp{\ifmmode\stackrel{<}{_{\sim}}\else$\stackrel{<}{_{\sim}}$\fi}
\def\gapp{\ifmmode\stackrel{>}{_{\sim}}\else$\stackrel{>}{_{\sim}}$\fi}

\def\spose#1{\hbox to 0pt{#1\hss}}
\def\approxlt{\mathrel{\spose{\lower 3pt\hbox{$\sim$}}
        \raise 2.0pt\hbox{$<$}}}
\def\approxgt{\mathrel{\spose{\lower 3pt\hbox{$\sim$}}
        \raise 2.0pt\hbox{$>$}}}

\def\lapp{\ifmmode\stackrel{<}{_{\sim}}\else$\stackrel{<}{_{\sim}}$\fi}
\def\gapp{\ifmmode\stackrel{>}{_{\sim}}\else$\stackrel{>}{_{\sim}}$\fi}

\slugcomment{Accepted for publication in The Astrophysical Journal Letters}

\begin{document}

\title{Revealing the Dusty Warm Absorber in \mcg6 with the Chandra HETG }
\author{
Julia C. Lee\altaffilmark{1},
Patrick M. Ogle\altaffilmark{1},
Claude R. Canizares\altaffilmark{1},
Herman L. Marshall\altaffilmark{1}, 
Norbert S. Schulz\altaffilmark{1}, 
Raquel Morales\altaffilmark{2},
Andrew C. Fabian\altaffilmark{2},
Kazushi Iwasawa\altaffilmark{2},
 }
\altaffiltext{1}{MIT Center for Space Research, 77 Massachusetts Ave., Cambridge, MA 02139.}
\altaffiltext{2}{Institute of Astronomy, Madingley Rd., Cambridge CB2 0HA  U.K.}

\medskip
\centerline{To appear in {\sc The Astrophysical Journal Letters}}

\begin{abstract}
We present detailed evidence for a warm absorber in the Seyfert~1
galaxy MCG--6-30-15 and dispute earlier claims for relativistic O line
emission.  The HETG spectra show numerous narrow, unresolved 
(FWHM $\approxlt$~200 $\kmps$) absorption lines  from a
wide range of ionization states of N, O, Mg, Ne, Si, S, Ar, and Fe.
The O~{\sc vii} edge and $1{\rm s}^2-1{\rm s}n{\rm p}$ resonance line
series to $n=9$  are clearly detected at rest in the AGN frame. We
attribute previous reports of an apparently  highly redshifted \ovii
edge  to the $1{\rm s}^2-1{\rm s}n{\rm p} \, ~(n > 5)$ \ovii resonance
lines, and a  neutral Fe~L   absorption complex.  The shape of the
Fe~L feature is nearly identical to that seen in the spectra of
several X-ray binaries, and in laboratory data.  The implied dust
column density agrees with that obtained from reddening studies, 
and gives the first direct  X-ray evidence for
dust embedded in a warm absorber. The O~{\sc viii} resonance lines and
weak edge are also detected, and the spectral rollover below
$\sim$~2~keV is explained by the superposition of numerous absorption
lines and edges. We identify, for the first time, a KLL resonance in
the O {\sc vi} photoabsorption cross section, giving a measure of the
O {\sc vi} column density.  The O~{\sc vii}~(f) emission detected at
the systemic velocity implies a covering fraction of $\sim 5$\% (depending
on the observed vs. time-averaged ionizing flux). Our
observations show that a dusty warm absorber model is not only adequate to
explain all the spectral features $\approxgt$ 0.48 keV ($\approxlt$ 26 \AA )
the data {\it require} it. This contradicts the interpretation of
Branduardi-Raymont \etal (2001) 
that this spectral region is dominated by highly relativistic line 
emission from the vicinity of the black hole.

\end{abstract}


\section{Introduction}
\label{sec:intro}
Recently, workers analyzing the \xmm RGS data of the luminous ($\rm L_X
\sim 10^{43} \ergps$) nearby ($z$=0.0078) Seyfert 1 galaxy \mcg6 and
the similar object Mrk~766 have proposed a radical alternative to the
warm absorber model as the origin of the spectral features in the soft
($<$ 2~keV) band (Branduardi-Raymont et al. 2001, hereafter BR2001).
It had been generally accepted that the features were imposed by
partially ionized absorbing material at $\approxgt$ parsec distances
from the black hole. The main signatures  are strong \ovii and \oviii
absorption edges (e.g. Fabian et al. 1994), which are nearly
ubiquitous in Seyfert~1 galaxies (e.g. Reynolds 1997,
George et al. 1998). However,  BR2001 propose that in \mcg6 and Mrk~766 the observed
spectral features are  soft X-ray emission lines from close to the
black hole  highly broadened by relativistic effects.  This scenario
is invoked to explain the apparent 16,000 $\kmps$ redshifted \ovii
edge without  associated resonance lines.

Our \chandra High Energy Transmission Grating (HETG) observation does
not support this interpretation.   We present strong evidence for
partially ionized absorbing material along the line of sight.   In
particular, we find absorption lines from a myriad of species  in a
range of ionization states which can explain the observed spectral
rollover at the energies of the warm absorber.  We demonstrate that
the  apparent redshifted \ovii edge can be  explained by the $1{\rm
s}^2-1{\rm s}n{\rm p}$ (for $n \approxgt 5$) \ovii resonance  lines
and a complex of Fe~L edges, plausibly from dust in the ionized
absorber. Here we concentrate on this evidence and leave more
comprehensive analyses to subsequent publications.

\section{Observations}
\label{sec:data}
MCG$-$6-30-15 was observed with the \chandra High Energy Transmission
Grating (HETG; Canizares \etal 2001) from 2000 April 5--6, and again from
2000 August 21--22. The total integration time was $\sim$~120~ks.
We reduce the spectral data from L1 (raw unfiltered event) files using IDL  processing
scripts which are similar to the standard CIAO processing. A more complete
description  is in Marshall, Schulz, \& Canizares (2001).
We restrict the event list to the nominal (0,2,3,4,6) grade set, and
remove event streaks in the S4 chip (these are not related to the
readout streak).   Event energies are
corrected for detector node-to-node gain variations.  The zeroth order 
position is determined and the $\pm 1$
order events are extracted for source and background regions.
Bad columns are eliminated, as well as data
which are affected by detector gaps.  We bin
the MEG (HEG) events at 0.01\AA\ (0.005\AA) intervals (an ACIS pixel)
for analysis.  The instrument effective area
is based on  pre-flight calibration data.

\section{Spectral Features}
The spectrum shows many features including numerous unresolved
absorption lines. The continuum exhibits a soft excess and a deficit
from 0.7~keV (17.7~\AA\,) to $\sim$ 2 keV (6.2~\AA\,), features long
attributed to a multi-zone warm absorber (e.g. Reynolds et al. 1995,
Otani et al. 1996). Our simultaneous \rxte data show the well-studied
power law and relativistically broadened iron line (e.g. Tanaka et
al. 1995), and will be reported separately.  We fit the \chandra
(0.45$-$0.67, 2.5$-$5~keV) MEG and (2$-$5, 6.5-8.~keV) HEG data
simultaneously obtaining $\Gamma \approx 1.84$  for the time-averaged
(120~ks) spectrum.  The MEG energies $0.7-2.5$~keV are   excluded from
our fits in order to mitigate the effect of the excess absorption in
that energy range; the Fe~K region is also excluded.   (This is
consistent with the best fit \rxte $3-10$~keV $\Gamma = 1.92 \pm
0.04$.)  The source varied by $\sim$50\% during each observation.
Correlated continuum flux and line strength variations were noted by
Otani et al. (1996) with \asca. Accordingly, we separated our data into
`high' and `low' states arbitrarily defined as above and below our mean
2 $\rm ct \, s^{-1}$.  We concentrate our study of the warm absorber
on the high state because many of the oxygen resonance absorption
lines appear most prominent during this state.
 
\subsection{Evidence for warm absorption}
\begin{figure*}
\psfig{file=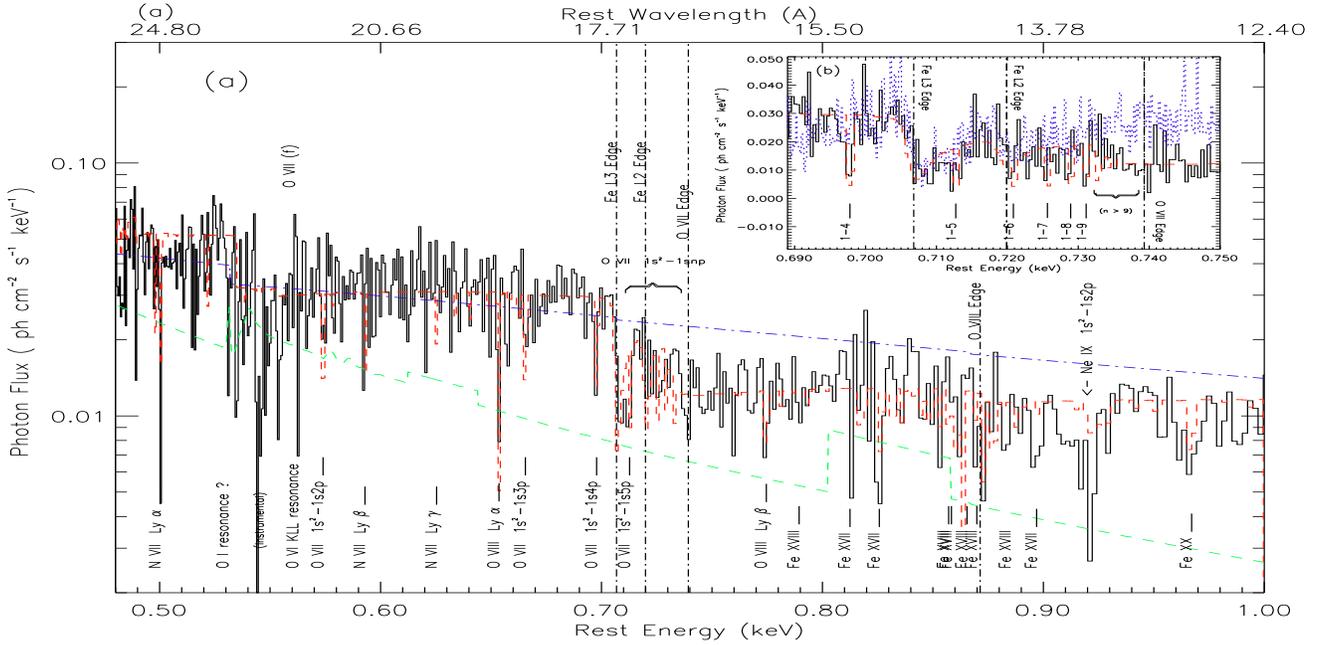,angle=90,width=1.00\textwidth,height=0.35\textheight}
\caption[h]{ (a) The \mcg6 spectrum (de-redshifted)  superposed on a
dusty warm absorber model (dashed;red), and  Galactic absorbed power-law
(dash-dots;blue). Plotted in dashed green is the approximate value of the $1 \sigma$
uncertainty; discrete jumps are due to regions in which the effective
area changes as a result of chip gaps.
Many of the lines predicted  by the model appear in
the MEG data. The dusty warm absorber model is binned to the same resolution
as the data (0.03~\AA\,).  The model includes the $n = 2-\infty$ \ovii
resonance series described in the text. The locations of the observed
and/or expected lines are marked.  (b)-inset The spectrum of Cygnus X1
(dotted; blue) and the  \ovii resonance series for n $>$ 3  (red) over-plotted on
\mcg6 (black) in the \ovii edge \& Fe~L region.  (The Cyg~X1 data have
been renormalized and exponentially scaled to the Fe~L3 edge depth of \mcg6, which
is de-redshifted).  }
\label{fig-waregion}
\end{figure*}

A strong testament to the existence of the warm absorber is the myriad
of ionized species in the 0.5--1~keV range, and in particular resonance
lines from N~{\sc vii}, O~({\sc vii,viii}), Ne~{(\sc ix,x)}, Fe~{\sc
xvii} to Fe~{\sc xxi} (Fig.~\ref{fig-waregion}a).  There are also many
absorption lines of higher ionization states of Fe up to Fe~{\sc
xxiii},  Mg, Si, S, Ar, and possibly Ca dispersed through the 0.9--5~keV
bandpass.  These lines, and the associated edges can plausibly account
for the observed spectral `rollover' $\approxlt$2 keV.

Of particular note are the resonance lines \ovii $1{\rm s}^2-1{\rm
s}n{\rm p}$ and \oviii~Ly$\alpha$, Ly$\beta$.  We detect \ovii $1{\rm
s}^2-1{\rm s}n{\rm p}$ up to $n=9$ with confidence levels from 
95\% to 99.9\%. These lines, like all the absorption lines discussed
in this paper are unresolved (implying FWHM $\approxlt$~200 $\kmps$), 
and at the expected wavelengths
to within $\approxlt 200 \kmps $ ($<$ 0.01~\AA\,) which is within
the HETG calibration uncertainty. The equivalent widths in the high
state for $n$=2, 4, 6--9 are respectively $\rm 17 \pm 9 \, (0.45 \,eV), 13 \pm 4 \, (0.51 \,eV),
 12 \pm 4, 11 \pm 4, 12 \pm 4, 13 \pm 4$~m\AA\ (the $n=$3,5 lines are confused
with the N~{\sc vii}
edge and the Fe~L3 feature, respectively). 
Clearly these lines are on the flat part of the curve of growth.
 We performed a curve-of-growth analysis for trial sets  of the parameters
turbulent velocity $b$ and column $N_{\rm OVII}$ (using the program
described in Pettini et al. 1983, Mar \& Bailey 1995) and find the 
minimum column density
implied by the absorption lines is $N_{O VII} \approxgt 7 \times
10^{17}  \psqcm$ ($b \sim 100 \kmps$) which requires an optical depth
\oviitau $> 0.2$  at the \ovii edge (Daltabuit \& Cox  1972).  Indeed,
Fig.~\ref{fig-waregion}a shows an edge  at  0.74~keV (16.8~\AA\,),
the energy expected for \ovii at rest in the AGN frame.  The drop
across the edge implies \oviitau $\sim 0.6-0.8$, or $N_{\rm OVII}
\approx 2.5 \times 10^{18} \psqcm$, which can account for the
observed strengths of the \ovii absorption series for $b \sim 100
\kmps$, consistent with the lines being unresolved in our data.
 The \oviii edge is not strong in our data (\oviiitau $\approx
0.15 \pm 0.10$).  and is complicated by a complex of nearby lines
from Fe~{\sc xvii} and Fe~{\sc xviii} species as well as the Ne~{\sc ix} 
$\rm 1s^2-1s2p$ resonance feature (e.g. Nicastro et al. 1999) shown in
Fig.~\ref{fig-waregion}a.  We find consistency between the \oviii
Ly$\alpha$ and Ly$\beta$ equivalent widths (respectively $23 \pm
5$~m\AA\,[0.79~eV] and $19 \pm 5$~m\AA\,[0.92~eV]),  and \oviii edge
for \oviiitau $\sim 0.1$, or $N_{\rm OVIII} \sim 1 \times 10^{18}
\psqcm$, and $b \sim 100 \kmps$.

We now investigate the strong drop at $\sim$~0.7~keV (if this is the
\ovii edge, then its redshift is $\sim 16,000 \kmps$ - e.g. BR2001),
and spectral features which  govern the \ovii complex of lines by
computing line strengths for the \ovii ($2 \le n < \infty$) resonance
series.   We use the line list of Verner et al. (1996) for the $n \le
10$  oscillator strengths and wavelengths, and extrapolate using the
hydrogenic approximation $f_{1n} \propto 1/n^3$, for $n > 10$.  We
take the observed \oviitau $\sim$~0.7, and generate the theoretical
spectrum shown up close in  Fig.~\ref{fig-waregion}b
(for $b$ = 100\kmps).  It can be seen that the apparent redshift of
the \ovii edge is partially explained by the \ovii $n > 5$ resonance
series.  We propose that both the $\sim 2500 \kmps$ wide feature and
residual absorption which cannot  be explained by the \ovii resonance
series are due to the L~edges of neutral iron.  The Fe~L3 edge at
0.707~keV (17.5~\AA\,) and the associated Fe~L2 edge seen in X-ray
binaries (e.g. X0614+011, Paerels et al., 2001; Cyg~X1, Schulz et al.,
2001; for laboratory data, see Crocombette \& Pollak 1995)  show structure
similar to that seen here.  To illustrate, Fig~\ref{fig-waregion}b
displays  the MEG spectra of \mcg6 (de-redshifted)  and Cyg~X1
(suitably scaled) in this region.  The agreement is remarkable, with
most of the differences accounted for by the \ovii edge and series
absorption.   The drop at 0.707~keV implies a neutral Fe column
density of  $\sim 4 \times 10^{17} \psqcm$ or $N_H  \sim 4 \times 10^{21}
\psqcm$ for solar abundances, but as noted in \S4, 
it is plausible to attribute the Fe to dust in the warm absorber, in
which case most of the remaining elements are ionized.  One does expect
other features, the most prominent being O (Cyg X-1 also shows weak Si and
Mg features which would not be detectable in our spectrum).  Using the dust
in Galactic halo clouds as a guide (Sembach \& Savage 1996, Savage \& Sembach 1996, argue for
a population of oxides and perhaps pure iron grains) suggests O/Fe $\sim 1-4$
(cf ., Snow \& Witt 1996).  An associated O I edge at the AGN redshift may
be present in our spectrum but is confused by the  instrumental O {\sc i } 
edge which has not yet been fully calibrated. (BR2001 report an
unspecified excess O absorption.)

We attribute the unresolved absorption line at $22.043 \pm
0.004$~\AA\, (0.562~keV; Fig.~\ref{fig-waregion}a), detected
at $\gg 99.9$\% confidence in the full observation, to the strongest
KLL resonance (photoexcitation to a
doubly excited state followed by auto-ionization) in the
photoabsorption cross section for O~{\sc vi} as recently calculated by
Pradhan (2000). The MEG easily separates this feature (which is itself
an unresolved doublet) from the nearby \ovii forbidden line at
22.09~\AA\,(561~eV) discussed below, giving an O~{\sc vi} equivalent
width of $29 \pm 8$ ~m\AA\,(0.74~eV).  Using Pradhan's effective
oscillator strength of 0.58 for this feature (assuming no
radiative damping and  $b \sim 100 \kmps$) gives $N_{O VI}\sim 3 \times 10^{17} \psqcm$.
(An associated O~{\sc vi} edge at 0.698~keV [17.76~\AA\,] is too weak
to be seen in our data.) 

We have made a first attempt at modeling the spectrum of \mcg6. While
not definitive, our model does explain the overall shape and most of
the absorption features.  In addition to the $\Gamma \sim 1.9$
power-law, we require a soft component and absorption by two ionized 
zones and neutrals in dust.  At least two zones are needed to explain
the wide range in ionization states that we observe from the absorption
species. (Multiple zones were also previously suggested for \mcg6
by Reynolds et al. 1995, Otani et al., 1996, Morales, Fabian, \& Reynolds 2000.)
In addition to the two ionized zones,
we add an extra column of $\rm 3.5 \times 10^{17} cm^{-2}$  of $\rm FeO_2$. 
We used the FeL cross sections provided by 
Kortright \& Kim (2000), and O I cross section from Henke et al. (1982).
 A neutral Fe:O ratio of 1:2  is 
supported by the data, and is consistent with dust composed of Fe oxides or 
olivine ($\rm Fe_2SiO_4$).  The Si edge would not be detectable in our data.
We did not attempt to model the O~{\sc vi} component.
We calculate ionization fractions and the absorption
spectrum of Fig.~\ref{fig-waregion} for the two ionized zones
using the photoionization code {\sc cloudy} (Ferland
1998), the complete \ovii resonance series, and the Verner et
al. (1996) resonance line list, for the relevant species.  The input
parameters to {\sc cloudy} assume a Galactic absorbed power-law
ionizing continuum with \xtegammamcg6, luminosity $10^{43} \, \ergps$,
and $b = 100 \kmps$.  
Emission from reconbination or resonance scattering is ignored (see below).
The low and high
ionization zones are respectively ($\rm log \xi$=0.7; $\rm log N_H$ =
21.7), and ($\rm log \xi$=2.5; $\rm log N_H$ = 22.5), where $\xi =
L/nR^2$.  The model includes a 10\% scattered continuum.

The multi-component warm absorber introduces an accumulation of
absorption lines and  edges (in particular various stages of Fe, Ne,
Mg) which causes the spectrum to roll over below 2~keV.  A soft excess
is required to compensate for this absorption at energies $<$~0.7~keV.
If it were not for the absorption, a single power-law with Galactic
absorption could roughly account for the  flux below 0.7~keV and above
2~keV.  Our data cannot distinguish between a power-law of $\Gamma
\sim 2.5$ or a 0.13~keV thermal black body for the soft excess.
(However, a 0.13~keV blackbody would imply a mass for \mcg6 $<< 10^6
\rm M_\odot$, which is implausible).


\subsection{\ovii (f) emission \& possible filling-in of absorption lines}
We detect the \ovii $1s2p\,$$^3S-1s^2\,$$^1S$ (forbidden) emission
line at $\lambda = 22.09 \pm 0.005$~\AA\, (561~eV) with $\approxgt$99.9\%
confidence and equivalent width $86 \pm 27$~m\AA\,  (2.2~eV) for
the combined observation.  This line is particularly strong in
photoionized plasmas and is unaffected  by resonance scattering, so it
is a good measure of the total rate of \ovii recombination (and
therefore photoionization) at the source.   We can obtain an estimate
for the  covering factor $f_{cov}$ of the \ovii absorber by comparing
the \ovii~(f) flux with the ionization rate along the line-of-sight
for \oviitau $\sim 0.7$.  Taking a branching ratio for \ovii~(f) of
$\sim 0.5$ per recombination (computed for $\sim 2 \times 10^5$~K,
following Porquet \& Dubau 2000, and Mewe \& Gronenschild 1981), we
find $f_{cov} \sim 0.05{\cal R}$, where ${\cal R}$ is the ratio of the
\ovii ionizing flux averaged over the light-travel time across the
warm absorber region to the flux averaged over our observation.

The presence of \ovii~(f) emission must be accompanied by  \ovii
$1s2p\,$$^1P-1s^2\,$$^1S$ resonance (r) line emission which  could
partially fill in the corresponding absorption line at $21.6$ ~\AA
\,(574~eV).  The (r)/(f) ratio from  recombination is $\sim 0.28$
(Porquet \& Dubau 2000) implying an equivalent width of $\sim
20$~m\AA \, in (r).    In addition to recombination, re-emission
(and possibly cascades) following resonance absorption can also fill
in some of the resonance lines. The actual amount of filling in from
such resonance scattering  is strongly dependent on geometry and
radiative transfer.  In any case, neglecting it strengthens our
arguments for lower limits to the ion column densities based on the
resonance absorption lines. From the \ovii~(f) line we estimate for
$\sim 2 \times 10^5$~K (using Nahar 1999 and Porquet \& Dubau 2000)
that the equivalent width of the radiative recombination (bound-free)
continuum from \ovii is $\sim 2$~eV with a likely width of $\approxgt
10$~eV, so it would not be prominent in our data.

\section{Discussion} \label{sec-discussion}
BR2001 concluded that their \xmm RGS observation of \mcg6 was
``physically and spectroscopically inconsistent'' with ``standard''
warm absorber models.  The thrust of their argument is the
inconsistency between the apparent highly redshifted \ovii  and \oviii
edges and the absence of corresponding resonance absorption lines,
which would be expected for any but the most contrived kinematic
models. The same conclusion is  reached for the similar spectrum of
Mkn 766. They then propose that the spectra of both sources be
explained by strong, highly  relativistically broadened Ly $\alpha$
emission lines of  H-like O, N and C from the near vicinity of a Kerr
black hole. This model implies that the rollover in the continuum below $\sim
2$~keV is due to a $\Gamma \sim 2.0$ to $\Gamma \sim 1.3$ break in the power-law.
The absorption lines they detect are attributed to low column density
absorbers with turbulent velocities $\sim 2000 \kmps$ FWHM; note that
such velocities would be easily resolved in our spectra, which have 
$\approxgt$ 3 times better resolution, but are not seen.

Our \chandra HETG spectrum of \mcg6 shows that a dusty warm-absorber model is
not only adequate to describe all the spectral 
features $\approxgt 0.48$ keV ($\approxlt 26$ \AA), the data {\it require} 
it.  Firstly, we observe an
abundance of highly-ionized resonance lines from a myriad of species
in a range of ionization states (from O~{\sc vi} to Si~{\sc xiii} to
Fe~{\sc xxiii}). The  complex of these absorption features and
associated edges will cause the  spectrum to roll over at $< 2$~keV,
but also to partially recover at lower energies. Secondly, we detect
the O~{\sc vii} $1{\rm s}^2-1{\rm s}n{\rm p} \, (n=2-9)$ resonance
absorption lines at the expected positions for absorbing material at
rest in the AGN frame.  The roughly comparable strengths of these
lines  indicate an O~{\sc vii} edge opacity of at least \oviitau $>$
0.2.  The value \oviitau $\sim 0.7$ indicated by the discontinuity at
the edge (at rest in the AGN frame) is fully consistent with the line
strengths for $b \sim 100 \kmps$. Thirdly, we show that the {\it
apparent} highly redshifted \ovii edge which prompted the BR2001
interpretation can be  explained in the context of a warm absorber. We
propose that it is partly due to the expected redward shift of the
absorption edge by the overlapping  \ovii $1{\rm s}^2-1{\rm s}n{\rm
p}$,  $n > 5$ series of  absorption lines, and partly to an Fe~L edge
complex from a significant column density of neutral Fe. The shape of
this complex is remarkably similar to that seen in the MEG data of Cyg
X-1 (Fig.~\ref{fig-waregion}b, Schulz \etal 2001).  A strong \oviii edge is not seen in our
data, however a small edge is fully consistent with the observed
\oviii Ly $\alpha$ and Ly $\beta$ absorption lines.  Previous reports
of a redshifted \oviii edge can be attributed to the series of Fe and
Ne lines near 0.87~keV (14.2~\AA) which could mimic a
broadened/shifted edge.  Our first attempts to model the full spectrum
also require an additional soft emission component, but this should be
broad (e.g., a steep power law or black body). Of course, we cannot
rule out some small contribution from relativistically broadened
emission lines as well, and our data cannot address the shape of the
spectrum below 0.48 keV. However, since we 
see no evidence for such line emission and can explain the great 
many features we do see
as coming from a dusty warm absorber, we conclude that 
the BR2001 interpretation of a spectrum dominated by relativistic
line emission is no longer tenable.

The most plausible explanation of the Fe~{\sc I}~L feature is
absorption by dust that is embedded in the partially ionized material.
[Grain survival in the dusty warm absorber of \mcg6 is addressed by
Reynolds et al. (1997).]
Otherwise, for cosmic abundances the implied neutral column density,
$N_H \sim 4 \times 10^{21} \psqcm$, would extinguish the X-ray spectrum at low
energies, contrary to the observations.  The presence of dust in the
warm absorber of \mcg6 was already identified by Reynolds et al. (1997)
based on the optical reddening of E(B--V) = 0.61--1.09, which implies
$N_H \approx  4-7 \times 10^{21} \psqcm$ (Cox 1999). Our independent
measurement based on the Fe~L absorption is consistent with this value.
Therefore the presence of neutral
absorption is not only explained, it is required by the observed reddening.
(Dusty warm absorbers have been suggested
for other AGN as well [e.g.  Brandt, Fabian, \& Pounds 1996, Reynolds
1997, Komossa 2000; see also Mathur 1994].)  This first detection of a spectral feature
attributed to the dust in an ionized absorber provides a new probe of
this component.  The shape of the Fe L features and the detection
or limits on other absorbing species may eventually constrain the chemical 
composition of the grains.
We note
that Mrk~766 (BR2001) also shows optical/UV extinction (Walter \& Fink
1993) suggesting that the above explanation for MCG--6-30-15 may also
apply.

We believe that our detection of an O {\sc vi} photoabsorption KLL
resonance is the first time such a feature has been detected in the
X-ray band (possibly including in the laboratory). It has two
important implications. First, it shows that even relatively low
ionization states previously detectable only in the UV can be studied
in detail in the X-ray band. Second, it underscores the importance of
having detailed calculations of the complex resonant structure in the
photoabsorption  cross sections of highly ionized species. Recent
calculations show great complexity (e.g. Nahar 1999, Zhang \& Pradhan
1999, Pradhan 2000) that has not yet been incorporated into the models
of warm absorbers.

The estimate of $\sim 5{\cal R}$\% placed on
$f_{cov}$  by the strength of the \ovii~(f) emission suggests that the
observed luminosity of the nucleus is $\approxgt$ its
average over the light-travel time across the \ovii region
($\sim$ years, see Otani \etal 1996)
else $f_{cov}$ would be implausibly small. 

These observations, and other recent work on AGN with \chandra and
\xmm (e.g. Kaastra \etal 2000, Kaspi \etal 2000, Ogle \etal 2000, Sako
\etal 2001a,b) illustrate the enormous power of high resolution
spectroscopy to probe the detailed physics of the ionized region in
AGN. They also show the complexity of these regions, as anticipated by
Netzer (1993), Nicastro (1999), Porquet \& Dubau (2000) and others.
For example, our data imply that the \mcg6 absorber spans a wide and
possibly continuous range of ionization, with embedded dust that will
introduce complex absorption features and also affect the state of the
plasma.  Resonance scattering and cascades are also likely to be
important, as is recombination line emission, as indicated by the
detection of \ovii~(f).  The absorbing region may also be a source of
thermal line emission, as seen in the HETG observation of NGC~4151
(Ogle \etal 2000).  We intend to pursue these issues in future
publications.

\section*{acknowledgments}
We  wish to thank Anil Pradhan
for useful conversations, Max Pettini for the use of  his
curve-of-growth program, Jeff Kortright, Kathryn Flanagan and 
Eric Gullikson for the latest experimental values for the atomic
 scattering factors in the FeL region.  We also thank
many of our colleagues in the MIT HETG/CXC group, with
special thanks to Dan Dewey.   We acknowledge the great efforts of the
many people who contributed to the \chandra program.  The work at MIT
was funded in part by contract SAO SV1-61010 and NASA contract
NAS8-39073.  ACF thanks the Royal Society for support.




\begin{references}

\reference{}
Brandt, W. N., Fabian, A. C., \& Pounds, K. A., 1996, MNRAS, 278, 326

\reference{}
Branduardi-Raymont, G., Sako, M., Kahn, S.M., Brinkman, A.C., Kaastra, J.S., Page, M.J., 2001, A\&A, 365, L140

\reference{}
Canizares, C.R., \etal 2001 (in preparation)

\reference{}
Crocombette, J.P., Pollak, M., Jollet, F., Thromat, N., \& Gautier-Soyer, M., 1995, Phys. Rev. B, 52, 3143

\reference{}
Cox, A., 1999, Allen's Astrophysical Quantities, Springer-Verlag, p. 197

\reference{}
Daltabuit E., \& Cox D.P., 1972, AJ, 177, 855

\reference{}
Fabian, A.C., et al., 1994, PASJ, 46, L59 

\reference{}
Ferland, G. J., Korista, K. T., Verner, D. A., Ferguson, J. W., Kingdon, J. B., Verner, E. M., 1998, PASP, 749, 761

\reference{}
George, I. M., Turner, T. J., Netzer, Hagai, Nandra, K., Mushotzky, R. F., Yaqoob, T., 1998, ApJS, 114, 73

\reference{}
Henke, B. L., Lee, P., Tanaka, T. J., Shimabukuro, R. L, Fujikawa, B. K., 1982, At. Data Nucl. Data Tables, 27, 1  

\reference{}
Kaastra, J. S., Mewe, R., Liedahl, D. A., Komossa, S., Brinkman, A. C. 2000 A\&A, 354, L83

\reference{}
Kaspi, S., Brandt, W.N., Netzer, H., Sambruna, R., Chartas, G., Garmire, G.P., Nousek, J.A. 2000 ApJL 535, L17

\reference{}
Komossa, S., 2000, in `ASCA/ROSAT Workshop on AGN and the X-ray Background', eds. T. Takahashi and H. Inoue, ISAS Report, 149, (astro-ph/0001263)

\reference{}
Kortright, J.B., \& Kim, S.-K., 2000, Phys. Rev. B, 62, 12216

\reference{}
Mar, D.P., \& Bailey, G., 1995, PASA, 12, 239


\reference{}
Marshall, H.L., Schulz N.S., Canizares C.R., 2001, ApJ, submitted

\reference{}
Mathur, S., 1994, ApJ., 431, L75


\reference{}
Morales, R.,  Fabian A.C.,, \& Reynolds C.S., MNRAS, 2000, 315, 149
 
\reference{}
Nahar, S. 1999 ApJ Sup 120, 131


\reference{}
Netzer, H. 1993 ApJ, 411, 594

\reference{}
Nicastro, F., Fiore, F., Perola, G.C., Elvis, M. 1999, ApJ 512, 184

\reference{}
Ogle, P.M., Marshall, H.L., Lee, J.C., Canizares, C.R., 2000, ApJ, 545, L81 

\reference{}
Otani, C., et al., 1996, PASJ, 48, 211

\reference{}
Mewe, R. \& Gronenschild, E.H.B.M., 1981, A\&A Supp, 45, 11

\reference{}
Paerels, F., Brinkman, A.C., van der Meer R.I.J., et al., 2001, ApJ, 546, 338 

\reference{}
Pettini, M., Hunstead, R.W., Murdoch, H.S., Blades, J.C., 1983, 273, 436

\reference{}
Porquet, D. \& Dubau, J. 2000 A\&A Sup 143, 495

\reference{}
Pradhan, A. 2000, ApJL, in press (astro-ph/0010255)

\reference{}
Reynolds, C.S., 1997, MNRAS, 286, 513

\reference{}
Reynolds, C. S., Fabian, A. C., Nandra, K., Inoue, H., Kunieda, H., Iwasawa, K., 1995, MNRAS, 277, 901

\reference{}
Reynolds C.S., Ward, M.J., Fabian, A.C. \& Celotti, A. 1997, MNRAS, 291, 403

\reference{}
Sako, M., Kahn, S.M., Behar, E., \etal., 2001, A\&A, 365, L168

\reference{}
Sako, M., Kahn, S.M., Paerels, F., \& Liedahl, D.A. 2001, ApJ., 543, L115

\reference{}
Schulz, N.S. \etal 2001 (in preparation)

\reference{}
Savage, B.D., \& Sembach, K.R., 1996, Ann. Rev Astron \& Astroph., 34, 279

\reference{}
Sembach, K.R., \& Savage, B.D., 1996, 457, 211

\reference{}
Snow T.P., \& Witt A.N., 1996, ApJ, 468, L65 

\reference{}
Tanaka Y., \etal, 1995, Nat., 375, 659

\reference{}
Verner, D.A., Verner, E.M., Ferland, G.J., 1996, BAAS, 188, 5418

\reference{}
Walter, R. \& Fink H.H., 1993, A\&A, 274, 105 

\reference{}
Zhang, H.L., \&  Pradhan, A.K., 1999, MNRAS, 313, 13

\end{references}
\end{document}